# Towards more reasonable identifications of the symmetries in noisy digital images from periodic and aperiodic crystals

P. Moeck

*Abstract*—A geometric form of information theory allows for reasonable, i.e. probabilistic, evidence-ranking based, and generalized noise-level dependent, classifications of the crystallographic and quasicrystallographic symmetries in noisy digital images. Such classifications are based solely on the image pixel intensity values, justifiable assumptions about the aggregate distribution of generalized noise in the images, asymptotic extrapolations to zero-noise images, and rational symmetry model selections with maximized predictive accuracy in the presence of both symmetry-inclusion relations and pseudo-symmetries. Preferring a well developed geometric form of information theory over a theoretically possible geometric-Bayesian approach for these classifications is the only subjective choice made. Using digital data planes and assuming approximately Gaussian distributed generalized noise, reasonable crystallographic and quasicrystallographic symmetry classifications can be made for noisy images from both scanning probe and transmission electron microscopes. A binary type classification of structurally very similar materials into either a quasicrystal or one of its rational/crystalline approximants based on the approximate point symmetries in their noisy digital images is proposed here for the first time.

## I. INTRODUCTION AND RESEARCH PROBLEM STATEMENT

Two of the three Nobel Prize winners in Chemistry of the year 2017 and 26 fellow experts in the field of electron microscopy (EM) as employed in structural biology concluded in the year 2012 with respect to the state-of-the-art of that field: *"The nature of the maps that are computed from 2D EM images depends principally on the symmetry of the objects being examined. Thus, there are maps for 2D crystal structures that can be obtained using either crystallographic methods (Henderson et al., 1986) or single-particle approaches (Frank et al., 1988) ... A satisfactory validation method does not yet exist and its development remains an open research problem."* [1]. These authors made the recommendation: *"A program is required to read in the density map, recognize any point group, helical or translational symmetry, and print out the point group or helical or space group symmetry, and the precise origin"* [1].

This author wholeheartedly agrees that such a program is needed and has in recent years developed efficient and subjective threshold-free methods for its operation in two dimensions (2D) for plane symmetry groups [2-4], projected Bravais lattice types [5,6], as well as projected (2D) crystallographic [2,4] and quasicrystallographic Laue classes. There is already some computer code by the author's research group for reasonable classifications into plane symmetry groups [3] and 2D crystallographic Laue classes as well as into 2D Bravais lattice types. This code is, however, not yet integrated and lacks a user friendly graphical user interface.

The goal of this paper is to inform the reader about the author's recent methodical developments [2-6]. The point will be made that objectivity and more reasonableness is needed at this point in time to address the above-mentioned research problem satisfactory for 2D projections of ordinary and sub-periodic crystals (3D) as well as their surfaces. The paper will also briefly make the case for using crystallographic image processing techniques in scanning probe microscopy (based on reasonable crystallographic symmetry classifications).

A multitude of different solutions to the crystallographic symmetry classification problem has been proposed over several decades, but the vast majority of them are marred by subjective thresholds for the interpretation of the aggregate pixel intensity values of digital 2D images of crystal patterns. For a recent review of the whole field of crystallographic symmetry classifications of such patterns in both reciprocal/Fourier and direct space, see [7].

A conclusion of the EM in structural biology validation taskforce report of the year 2020 (by some of the same authors as in [1]) reads: *"... as currently practiced, the procedure is not sufficiently standardized: a number of different variables (e.g. ... threshold value for interpretation) can substantially impact the outcome. As a result, different expert practitioners can arrive at different resolution estimates for the same level of map details"* [8].

The direct quotes above were selected for their effectiveness and conciseness in summing up what leading EM-based structural biologists think with respect to scientific problems that need to be overcome.

## II. AWARENESS OF THE SUBJECTIVITY THAT COMES WITH THRESHOLDS FOR IMAGE INTERPRETATIONS IN DIFFERENT RESEARCH COMMUNITIES

The above-given direct quote from [8] implies that as of the present time, structural biologists are not using the kind of program that was recommended to be developed in [1] and which would take the subjectivity out of their interpretation of experimental data. Many structural biologists who use the electron crystallography approach are,

Peter Moeck is with the Nano-Crystallography Group, Department of Physics, Portland State University, Portland, OR 97207-0751, phone: 503-725-4227; fax: 503-725-2815; e-mail: pmoeck@pdx.edu.

however, well aware that having both more reasonable approaches to symmetry classifications than the traditionally used ones and computer programs that embrace those approaches would be a "game changer" in their field. Expert practitioners would then no longer have to use subjectively set symmetry thresholds in the interpretation of their EM data and electron density maps.

The implications of the crystallographic symmetry classification problem in 2D are particularly severe in the electron crystallography based structural biology community. This is because intrinsic membrane proteins form sub-periodic, i.e. only one unit cell thick, crystals that are highly sensitive to damage by the electron beam so that their digital transmission electron microscope (TEM) images can only have very poor signal to noise (S/N) ratios. The analysis of such proteins by cryo-electron microscopy has for that reason been called *"the "bête noire" for crystallographic analysis"* in a scientific profile paper on the three 2017 Nobel Prize laureates in Chemistry [9]. In addition, the contrast in their electron micrographs is very weak. It is for these reasons not surprising that members of that community are outspoken about the importance of making crystallographic symmetry classifications as objectively as possible [1,8].

Most of the inorganic crystalline materials that electron microscopists work on in materials science tolerate much higher electron doses so that good S/N ratios are obtained. Also, their electron micrographs have typically a much higher contrast. Similarly, scanning probe microscope (SPM) images typically have good S/N ratios and contrasts. It is, therefore, also no surprise that most materials scientists and scanning probe microscopists seem at present satisfied with the undeniable subjectivity of the traditional "gut feeling" based crystallographic symmetry classification methods.

To this author, all of those classifications are quite unreasonable due to their subjectivity [7] and especially because reasonable alternatives [2-6] became available in recent years. A statistically sound dealing with the well known symmetry inclusion relations is in those classifications negated by a researcher's gut feelings. Pseudo-symmetries are ignored although they can be very difficult to distinguish from their symmetry counterparts in the unavoidable presence of noise. Those classifications are, accordingly, bound to be misleading or wrong on multiple occasions. They are also presented definitively, i.e. are communicated with an ascribed probability of 100% (meaning certainty), which is factually an impossibility.

Only a few members of the SPM community have so far made crystallographic symmetry classifications for images from crystalline sample surfaces. Fourier filtering [10] is the preferred method to process raw SPM images, which is equivalent to averaging the pixel values over all unit cells in the image. Experimental SPM images from crystals could, however, also be processed crystallographically, provided one has made reasonable symmetry classifications first. This idea will be further expanded on in the penultimate section of this paper.

### III. GENERAL DATA PLANES AND GENERALIZED NOISE

When considered as digital 2D data planes of a general nature from which structural and physical information is to be extracted in the most effective manner and in the unavoidable presence of experimental noise, the specifics of the physical interactions that resulted in the digital images become unimportant. The same basic crystallographic symmetry classification approach works, therefore, for images from atomic and molecular resolution TEMs and various SPMs which achieve that kind of resolution.

Generalized noise includes all effects of unavoidably imperfect recordings of images including shot noise, all kinds of rounding effects and numerical approximations by any kind of image processing algorithm, all structural defects in crystalline and quasicrystalline real-world samples, as well as uneven negative staining in cryo-TEM. When there are many different sources of noise and the effects of none of these sources dominates, the resulting generalized noise can justifiably be assumed to be approximately Gaussian distributed (by the Lindeberg-Lévy, Lindeberg-Feller, and Lyapunov generalizations of the well known central limit theorem of statistics). This distribution is the precondition for the application of geometric Akaike Information Criteria [2-6,11].

The assumption that the generalized noise in digital images that are to be classified with respect to their crystallographic and quasicrystallographic symmetries is approximately Gaussian distributed is quite reasonable when there is no information beyond the individual images themselves available [12]. Additional information to eventually make the author's approaches [2-6] to crystallographic and quasicrystallographic symmetry classifications even more reasonable in the future can only come from experimental investigations of the statistical properties of noise for different SPMs and TEMs for a multitude of real world images. The author's research group is planning to undertake such studies, but will use data planes as representations of digital images for some time to come due to the inherent simplicity of that particular approach.

### IV. GEOMETRIC AKAIKE WEIGHTS AND CRYSTALLOGRAPHICALLY CONSISTENT SYMMETRY CLASSIFICATIONS

An Akaike weight is the probability that the corresponding mathematical model for observational or experimental data is the Kullback-Leibler (KL) best model in a set of alternative models for this data [13]. It can also be interpreted as the quantified proportion of total evidence in support of a particular model within the selected model set as being the KL-best model.

Non-geometric Akaike weights allow for evidence rankings of mathematical models with respect to their parsimonious predictive capability in representing data in the limit of an infinite number of observations or experimental runs. The KL-best model minimizes the divergence between the data and its multiple model representations. The picking of the mathematical model that has the highest probability of minimizing the expected KL divergence as the best representation of observational or experimental data is a statistically sound model selection process. This is equivalent to the maximization of the data's relative Shannon entropy (expected information content) under the constraints of the alternative mathematical models. Akaike weights are also Bayesian posterior model probabilities on the basis of "*savvy priors*" [13].

Probabilistic symmetry classifications are made on geometric data so that geometric Akaike weights (G-AW) are required. They have been introduced in [2] and follow the same logic as the Akaike weights of the likelihood statistics of non-geometric observational or experimental data [13]. The well known Akaike Information Criterion (AIC) of frequentist observation based likelihood statistics needed to be replaced with a geometric AIC [11]. The mathematical model set is for crystallographic symmetries precisely defined [14].

The assumption is usually made that the image that is to be classified features in the limit of vanishing noise either site symmetries higher than the identity operation or at least a set of glide lines. Cases beyond the validity of this assumption are revealed by comparatively high sums of squared residuals for *all* symmetry models. A point/site pseudo-symmetry of an image that features only translation symmetry may be mistaken for a genuine symmetry when noise levels are relatively high.

Ratios of geometric AIC values of non-disjoint symmetry models [2,4] of the image data are very useful for the identification of pseudo-symmetries. A need to estimate the generalized noise level no longer exists when such ratios are considered [11]. The most probable joint symmetry classification of a more or less 2D periodic image into a plane symmetry group and Laue class needs not only to have high G-AW values for both symmetry types, but also needs to be crystallographically consistent. See [14] for what this consistency entails. Translational pseudo-symmetries, metric specializations, and 2D Bravais lattice types are discussed in [15].

## V. Binary type classification: Crystalline or quasicrystalline?

Efficient classifications into crystallographic [2] and quasicrystallographic Laue classes are made in Fourier/reciprocal space. The former Laue classes are a subset of the crystallographic point symmetry groups. All 2D Laue classes are point symmetries that feature two-fold rotation points and are the possible symmetries of the amplitude maps (or power spectra) of discrete Fourier transforms of direct space image data. In reciprocal 2D space, the center of the amplitude map of a discrete Fourier transform coincides with a two-fold rotation point regardless of such points actually being present (to some approximation) in the noisy direct space 2D image. This is so because the Fourier transform is centrosymmetric.

Quasicrystals feature point symmetries that do not obey the restrictions that are imposed by the translation symmetry of ordinary crystals. The quasicrystallographic Laue classes capture the point symmetries of quasicrystals up to the consequence of the Fourier transform creating a two-fold rotation point that is added to the prevailing point symmetry group. All of the 2D quasicrystallographic Laue classes are in symmetry inclusion relationships with the 2D crystallographic Laue classes.

For all quasicrystals, there are also quasicrystal approximants that are rather similar in their atomic structure, but possess crystallographic Laue classes due to the translation symmetry restrictions on ordinary crystals. As a consequence of the structural similarities to their quasicrystal counterpart, the crystalline approximants feature prominent pseudo-symmetries at the point symmetry level that are non-crystallographic.

Reasonable classifications [2,3] of an image into a Laue class using the amplitude map of a discrete Fourier transform that are based on informatiton theory [11,13] allow one to distinguish if an experimental image was recorded from a quasicrystal or from one of its crystalline approximants. When one classifies the point symmetry in the amplitude maps of discrete Fourier transforms with information theory based direct space methods, no indexing of the structure bearing Fourier coefficients is required for making such binary crystal type distinctions. Following ideas outlined in [16], one may alternatively use an approximate 2D indexing scheme.

## VI. How scanning probe microscopy stands to benefit from reasonable crystallographic symmetry classifications

Electron crystallographers in both structural biology and materials science use the estimated plane symmetry groups that their images probably possess in the process of crystallographic image processing as part of an elucidation of the structure of their crystals. Scanning probe microscopists can enhance their measurements of the spatial distribution of physical properties that emerge from the specifics of the underlying crystalline structure by doing crystallographic image processing as well.

There is precedence for doing this with the (reasonable) information theory based approach recently [17] and earlier in the traditional (subjective) manner [18-25]. Crystallographic image processing performs an averaging over all asymmetric units of all unit cells, significantly enhancing the signal to noise ratio of an image with respect to what is obtainable by standard [10] Fourier filtering alone.

When the experimental image has been classified into the plane symmetry group that most likely underlies the aggregate of all pixel intensity values (in the KL-best geometric model sense given all symmetry alternatives, justifiable assumptions, and the noisy image itself), asymmetric unit averaging enables much better extractions of the physical signal from symmetrized SPM image data.

Crystallographic image processing of SPM images from crystals can also be understood as posterior (software-based) symmetrizing/sharpening of the experimental scan-ning probe tip [22-25]. This results in an increased microscope resolution and allows also for the extraction of the prevailing point spread function [18,21,22] of the microscope. This point spread function can be determined on a crystalline calibration sample and its inverse can be used to deconvolute SPM images from other samples that were recorded with the same tip under essentially the same experimental conditions and microscope settings [21].

## VII. CLOSING REMARKS

Note finally that the author's research group has not taken up the development of reasonable methods (and associated computer programs) for the classifications of single biological 3D particles into point symmetries on the basis of their projections into 2D, as was implicitly stated as an unsolved research problem in the direct quotes from [1]. The author has no plans to do so in the future as he is an applied crystallographer and materials scientist.

There is no translation periodicity restriction on the point symmetries that these particles can possess in 2D so that symmetry classifications become in the unavoidable presence of imaging noise a nondeterministic polynomial time hard (NP-hard) problem [26]. There might, however, be other reasonable restrictions, e.g. rotation points that correspond to low Fibonacci numbers, so that the NP-hard aspects of the classification problem could be bypassed.

The author is also not interested in working on more reasonable classification methods for biological particles with helical symmetry on the basis of Fourier-Bessel methods. He leaves solutions to the corresponding *"open research problem"* [1] to others.